\documentstyle[12pt]{book}
\begin{document}

\sloppy
\raggedbottom

\chapter* 
{RANDOMNESS IN ARITHMETIC AND THE DECLINE AND FALL OF REDUCTIONISM IN PURE
MATHEMATICS}
\markright
{The Decline and Fall of Reductionism in Pure Mathematics}
\addcontentsline{toc}{chapter}
{The decline and fall of reductionism}

\section*{IBM Research Report RC-18532\\November 1992}
\section*{}
\section*{G. J. Chaitin}
\section*{}

{\it
Lecture given Thursday 22 October 1992
at a Mathematics -- Computer Science Colloquium
at the University of New Mexico.
The lecture was videotaped; this is an edited transcript.
}

\section*{1. Hilbert on the axiomatic method}

Last month I was a speaker at a symposium on reductionism at Cambridge
University
where Turing did his work.  I'd like to repeat the talk I gave there and
explain how
my work continues and extends Turing's.
Two previous speakers had said bad things about David Hilbert.
So I started by saying that
in spite of what you might have heard in some of the previous lectures, Hilbert
was not
a twit!

Hilbert's idea is the culmination of two
thousand years of mathematical tradition going back to Euclid's axiomatic
treatment of geometry, going back to Leibniz's dream of a symbolic logic
and Russell and Whitehead's monumental {\it Principia Mathematica.}
Hilbert's dream was to once and for all clarify the methods of mathematical
reasoning.  Hilbert wanted to formulate a formal axiomatic system which
would encompass all of mathematics.
\[
\begin{array}{l}
   \mbox{\bf Formal Axiomatic System} \\
   \longrightarrow \\
   \longrightarrow \\
   \longrightarrow
\end{array}
\]

Hilbert emphasized a number of
key properties that such a formal axiomatic system should have.  It's
like a computer programming language.  It's a precise
statement about the methods of reasoning, the postulates
and the methods of inference that we accept as mathematicians.
Furthermore Hilbert stipulated that the formal axiomatic system
encompassing all of mathematics that he wanted to construct
should be ``consistent'' and it should be ``complete.''
\[
\begin{array}{l}
   \mbox{\bf Formal Axiomatic System } \\
   \mbox{$\longrightarrow$ consistent} \\
   \mbox{$\longrightarrow$ complete  } \\
   \mbox{$\longrightarrow$           }
\end{array}
\]

Consistent means that you shouldn't be able
to prove an assertion and the contrary of the assertion.
\[
\begin{array}{l}
   \mbox{\bf Formal Axiomatic System } \\
   \mbox{$\longrightarrow$ consistent $A$ $\neg A$} \\
   \mbox{$\longrightarrow$ complete  } \\
   \mbox{$\longrightarrow$           }
\end{array}
\]
You shouldn't be able to prove $A$ and not $A$.
That would be very embarrassing.

Complete means that if
you make a meaningful assertion you should be able to
settle it one way or the other.  It means that either $A$ or not $A$
should be a theorem,
should be provable from the axioms using
the rules of inference in the formal axiomatic system.
\[
\begin{array}{l}
   \mbox{\bf Formal Axiomatic System } \\
   \mbox{$\longrightarrow$ consistent $A$ $\neg A$} \\
   \mbox{$\longrightarrow$ complete   $A$ $\neg A$} \\
   \mbox{$\longrightarrow$           }
\end{array}
\]
Consider a meaningful assertion $A$ and its contrary not $A$.
Exactly one of the two should be provable
if the formal axiomatic system is consistent and complete.

A formal axiomatic system is like a programming language.
There's an alphabet and rules of grammar, in other words,
a formal syntax.  It's a kind of thing that we are
familiar with now.
Look back at Russell and Whitehead's three
enormous volumes full of symbols and you'll feel you're looking
at a large computer program in some incomprehensible programming
language.

Now there's a very surprising fact.
Consistent and complete means only truth and all the truth.
They seem like reasonable requirements.  There's a funny consequence,
though, having to do with something called the decision problem.
In German it's the Entscheidungsproblem.
\[
\begin{array}{l}
   \mbox{\bf Formal Axiomatic System } \\
   \mbox{$\longrightarrow$ consistent $A$ $\neg A$} \\
   \mbox{$\longrightarrow$ complete   $A$ $\neg A$} \\
   \mbox{$\longrightarrow$ decision problem       } \\
\end{array}
\]

Hilbert ascribed a great deal of importance to the decision problem.
\[
\begin{array}{l}
   \mbox{\bf HILBERT                 } \\
   \mbox{\bf Formal Axiomatic System } \\
   \mbox{$\longrightarrow$ consistent $A$ $\neg A$} \\
   \mbox{$\longrightarrow$ complete   $A$ $\neg A$} \\
   \mbox{$\longrightarrow$ decision problem       } \\
\end{array}
\]

Solving the decision problem for a
formal axiomatic system is giving an algorithm
that enables you to decide whether any given meaningful assertion
is a theorem or not.
A solution of the decision problem is called a decision procedure.
\[
\begin{array}{l}
   \mbox{\bf HILBERT                 } \\
   \mbox{\bf Formal Axiomatic System } \\
   \mbox{$\longrightarrow$ consistent $A$ $\neg A$} \\
   \mbox{$\longrightarrow$ complete   $A$ $\neg A$} \\
   \mbox{$\longrightarrow$ decision procedure     } \\
\end{array}
\]

This sounds weird.  The formal axiomatic system
that Hilbert wanted to construct would have included all of
mathematics: elementary arithmetic, calculus, algebra, everything.
If there's a decision procedure, then
mathematicians are out of work.
This algorithm, this mechanical procedure, can check whether
something is a theorem or not, can check whether it's true
or not.  So to require that there be a decision
procedure for this formal axiomatic system sounds like you're asking for a lot.

However it's very easy to see
that if it's consistent and it's complete that
implies that there must be a decision procedure.
Here's how you do it.  You have a formal language with a finite
alphabet and a grammar.  And Hilbert emphasized that the whole point of a
formal axiomatic system
is that there must be a mechanical procedure
for checking whether a purported proof is correct or not, whether it obeys
the rules or not.
That's the notion that mathematical truth should be objective so that
everyone can agree whether a proof follows the rules or not.

So if that's
the case you run through all possible proofs in size order, and look
at all sequences of symbols from the alphabet one character long, two,
three, four, a thousand, a thousand and one\ldots\ a hundred thousand
characters
long.  You apply the mechanical procedure which is the essence of
the formal axiomatic system, to check whether each proof is valid.  Most of the
time, of course, it'll be nonsense, it'll be ungrammatical.  But you'll
eventually
find every possible proof.
It's like a million monkeys typing away.  You'll find every possible proof,
though only
in principle of course.  The number grows exponentially and this is something
that you couldn't do in practice.  You'd never get to proofs that are one page
long.

But in principle you could run through all possible
proofs, check which ones are valid, see what they prove, and that
way you can systematically find all theorems.  In other words, there is an
algorithm,
a mechanical procedure, for generating one by one every theorem
that can be demonstrated in a formal axiomatic system.
So if for every meaningful assertion within the system,
either the assertion is a theorem or its contrary is a theorem,
only one of them, then you get a decision procedure. To see whether an
assertion is a theorem or not you just run through all
possible proofs until you find the assertion coming out as a theorem
or you prove the contrary assertion.

So it seems that Hilbert actually believed that he was going to
solve once and for all, all mathematical problems.
It sounds amazing, but apparently he did.
He believed that he would be able
to set down a consistent and complete formal axiomatic system for all
of mathematics and from it obtain a decision procedure for all of mathematics.
This is just following the formal, axiomatic tradition in mathematics.

But I'm sure he didn't think that it would be a practical decision procedure.
The one I've outlined would only work in principle.  It's
exponentially slow, it's terribly slow!  Totally impractical.
But the idea was that if all mathematicians could agree whether a proof is
correct and be consistent and complete, in principle that would give
a decision procedure for automatically solving any mathematical problem.
This was Hilbert's magnificent dream, and it was to be the culmination
of Euclid and Leibniz, and Boole and Peano, and Russell and Whitehead.

Of course the only
problem with this inspiring project is that it turned out to be impossible!

\section*{2. G\"odel, Turing and Cantor's diagonal argument}

Hilbert is indeed inspiring.  His famous lecture in the year 1900 is
a call to arms to mathematicians to solve a list of twenty-three difficult
problems.
As a young kid becoming a
mathematician you read that list of twenty-three problems and Hilbert is
saying that there is no limit to what mathematicians can do.
We can solve a problem if we are clever enough and work at it long
enough.  He didn't believe that in principle there was any limit to what
mathematics
could achieve.

I think this is very inspiring.  So did John von Neumann.
When he was a young man he tried to carry through Hilbert's ambitious
program.  Because Hilbert couldn't quite get it all to work, in fact he
started off just with elementary number theory, 1, 2, 3, 4, 5, \ldots,
not even with real numbers at first.

And then in 1931 to everyone's great
surprise (including von Neumann's), G\"odel showed that it was impossible,
that it couldn't be done, as I'm sure you all know.
\[
   \mbox{{\bf G\"odel} 1931}
\]

This was the opposite of what everyone had expected.
Von Neumann said it never occurred to him that Hilbert's program
couldn't be carried out.  Von Neumann admired G\"odel enormously, and
helped him to get a permanent position at the Institute for Advanced Study.

What G\"odel showed was the following.  Suppose that
you have a formal axiomatic system dealing with elementary
number theory, with 1, 2, 3, 4, 5 and addition and multiplication.  And we'll
assume that
it's consistent, which is a minimum requirement---%
if you can prove false results it's really pretty bad.
What G\"odel showed was that if you assume that it's consistent,
then you can show that it's incomplete.
That was G\"odel's result, and the proof is very clever and involves
self-reference.  G\"odel was able to construct an assertion
about the whole numbers that says of itself that it's unprovable.
This was a tremendous shock.  G\"odel has to be admired
for his intellectual imagination; everyone else thought that Hilbert was right.

However I think that Turing's 1936 approach is better.
\[
\begin{array}{l}
   \mbox{{\bf G\"odel} 1931} \\
   \mbox{{\bf Turing}  1936}
\end{array}
\]
G\"odel's 1931 proof is very ingenious, it's a real tour de
force.  I have to confess that when I was a kid trying to understand it,
I could read it and follow it step by step but somehow
I couldn't ever really feel that I was grasping it.  Now Turing had a
completely
different approach.

Turing's approach I think it's fair to say is in some ways more
fundamental.  In fact, Turing did
more than G\"odel.  Turing not only got as a corollary G\"odel's result,
he showed that there could be no decision procedure.

You see, if you assume that you have a formal axiomatic system for
arithmetic and it's consistent, from G\"odel you know that it
can't be complete, but there still might be a decision procedure.  There
still might be a mechanical procedure which would enable you to decide if
a given assertion is true or not. 
That was left open by G\"odel, but Turing settled it.
The fact that there cannot be a decision procedure is
more fundamental and you get incompleteness as a corollary.

How did Turing do it?
I want to tell you how he did it because that's the springboard for my own
work.
The way he did it, and I'm sure all of you have heard about it, has to do with
something called the halting problem.
In fact if you go back to Turing's 1936 paper you will not find the words
``halting problem.''  But the idea is certainly there.

People also forget that Turing was talking about ``computable numbers.''
The title of his paper is
``On computable numbers, with an application to the Entscheidungsproblem.''
Everyone remembers that the halting problem is unsolvable and that comes
from that paper,
but not as many people remember that Turing was talking about
computable real numbers.  My work deals with computable and
dramatically uncomputable real numbers.  So I'd like to refresh your
memory how Turing's argument goes.

Turing's argument is really what destroys Hilbert's dream, and it's a simple
argument.  It's just Cantor's diagonal procedure (for those of you who
know what that is) applied to the computable real numbers.
That's it, that's the whole idea in a nutshell, and it's
enough to show that Hilbert's dream, the culmination of two
thousand years of what mathematicians thought mathematics was about,
is wrong.  So Turing's work is tremendously deep.

What is Turing's argument?  A real number,
you know $3.1415926\cdots$, is a length measured with arbitrary precision,
with an infinite number of digits.
And a computable real number said Turing is one for which there is a computer
program or algorithm for calculating the digits one by one.
For example, there are programs for $\pi$, and there are algorithms for
solutions of
algebraic equations with integer coefficients.
In fact most of the numbers that you actually find in analysis
are computable.
However they're the exception, if you know set theory, because the computable
reals are denumerable and the reals are nondenumerable (you don't have
to know what that means).  That's the essence of Turing's idea.

The idea is this.  You list all possible computer programs.
At that time there were no computer programs,
and Turing had to invent the Turing machine, which was a tremendous step
forward.
But now you just say, imagine writing a list with every possible computer
program.
\[
\begin{array}{llll}
\begin{array}[t]{l}
   p_1 \\
   p_2 \\
   p_3 \\
   p_4 \\
   p_5 \\
   p_6 \\
   \vdots
\end{array}
& & &
\begin{array}[t]{l}
   \mbox{{\bf G\"odel} 1931} \\
   \mbox{{\bf Turing}  1936}
\end{array}
\end{array}
\]

If you consider computer programs to be in binary, then it's natural to think
of a computer program as a natural number.  And next to each computer
program, the first one, the second one, the third one, write out the real
number that it computes if it computes a real (it may not).
But if it prints out an infinite number of digits, write them out.  So maybe
it's
3.1415926 and here you have another and another and another:
\[
\begin{array}{llll}
\begin{array}[t]{l}
   p_1 \;\; 3.1415926\cdots \\
   p_2 \;\; \cdots \\
   p_3 \;\; \cdots \\
   p_4 \;\; \cdots \\
   p_5 \;\; \cdots \\
   p_6 \;\; \cdots \\
   \vdots
\end{array}
& & &
\begin{array}[t]{l}
   \mbox{{\bf G\"odel} 1931} \\
   \mbox{{\bf Turing}  1936}
\end{array}
\end{array}
\]

So you make this list.  Maybe some of
these programs don't print out an infinite number of
digits, because they're programs that halt or that have an error in them
and explode.  But then there'll just be a blank line in the list.
\[
\begin{array}{llll}
\begin{array}[t]{l}
   p_1 \;\; 3.1415926\cdots \\
   p_2 \;\; \cdots \\
   p_3 \;\; \cdots \\
   p_4 \;\; \cdots \\
   p_5 \;\;        \\
   p_6 \;\; \cdots \\
   \vdots
\end{array}
& & &
\begin{array}[t]{l}
   \mbox{{\bf G\"odel} 1931} \\
   \mbox{{\bf Turing}  1936}
\end{array}
\end{array}
\]
It's not really important---let's forget about this possibility.

Following Cantor, Turing says go
down the diagonal and look at the first digit of the first number,
the second digit of the second, the third\ldots
\[
\begin{array}{ll}
\begin{array}[t]{l}
   p_1 \;\; -\!.\underline{d_{11}}d_{12}d_{13}d_{14}d_{15}d_{16}\cdots \\
   p_2 \;\; -\!.d_{21}\underline{d_{22}}d_{23}d_{24}d_{25}d_{26}\cdots \\
   p_3 \;\; -\!.d_{31}d_{32}\underline{d_{33}}d_{34}d_{35}d_{36}\cdots \\
   p_4 \;\; -\!.d_{41}d_{42}d_{43}\underline{d_{44}}d_{45}d_{46}\cdots \\
   p_5 \;\;                                                            \\
   p_6 \;\; -\!.d_{61}d_{62}d_{63}d_{64}d_{65}\underline{d_{66}}\cdots \\
   \vdots
\end{array}
&
\begin{array}[t]{l}
   \mbox{{\bf G\"odel} 1931} \\
   \mbox{{\bf Turing}  1936}
\end{array}
\end{array}
\]

Well actually it's the
digits after the decimal point.  So it's the first digit after the
decimal point of the the first number,
the second digit after the decimal point of the second,
the third digit of the third number, the fourth digit of the fourth,
the fifth digit of the fifth.  And it doesn't matter
if the fifth program doesn't put out a fifth digit, it really doesn't matter.

What you do is you change these digits.  Make them different.
Change every digit on the diagonal.  Put these changed digits together into a
new number
with a decimal point in front, a new real number.
That's Cantor's diagonal procedure.
So you have a digit which
you choose to be different from the first digit of the first number, the
second digit of the second, the third of the third, and you put these together
into one number.
\[
\begin{array}{ll}
\begin{array}[t]{l}
   p_1 \;\; -\!.\underline{d_{11}}d_{12}d_{13}d_{14}d_{15}d_{16}\cdots \\
   p_2 \;\; -\!.d_{21}\underline{d_{22}}d_{23}d_{24}d_{25}d_{26}\cdots \\
   p_3 \;\; -\!.d_{31}d_{32}\underline{d_{33}}d_{34}d_{35}d_{36}\cdots \\
   p_4 \;\; -\!.d_{41}d_{42}d_{43}\underline{d_{44}}d_{45}d_{46}\cdots \\
   p_5 \;\;                                                            \\
   p_6 \;\; -\!.d_{61}d_{62}d_{63}d_{64}d_{65}\underline{d_{66}}\cdots \\
   \vdots                                                              \\
.\neq\!d_{11}\neq\!d_{22}\neq\!d_{33}\neq\!d_{44}\neq\!d_{55}\neq\!d_{66}\cdots
\end{array}
&
\begin{array}[t]{l}
   \mbox{{\bf G\"odel} 1931} \\
   \mbox{{\bf Turing}  1936}
\end{array}
\end{array}
\]

This new number cannot be in the list because of the way it was
constructed.  Therefore it's an uncomputable real number.  How does Turing
go on from here to the halting problem?  Well, just ask
yourself {\bf why} can't you compute it?  I've explained how to
get this number and it looks like you could almost do it.
To compute the $N$th digit of this number, you get the
$N$th computer program (you can certainly do that) and then you start it
running
until it puts out an $N$th digit, and at that point you change it.
Well what's the problem?  That sounds easy.

The problem is, what happens if the $N$th computer program never puts out
an $N$th digit, and you sit there waiting?  And that's the halting problem---%
you cannot decide whether the $N$th computer program will ever put out an $N$th
digit!
This is how Turing got the unsolvability of the halting problem.
Because if you could solve the halting problem, then
you could decide if the $N$th computer program
ever puts out an $N$th digit.  And if you could do that then you could actually
carry
out Cantor's diagonal procedure and
compute a real number which has to differ from any computable
real.  That's Turing's original argument.

Why does this explode Hilbert's dream?  What has
Turing proved?  That there is no algorithm, no mechanical procedure, which will
decide
if the $N$th computer program ever outputs an $N$th digit.
Thus there can be no algorithm which will decide if a computer
program ever halts (finding the $N$ digit put out by the $N$th program is a
special case).
Well, what Hilbert wanted was a formal axiomatic system from
which all mathematical truth should follow, only mathematical truth, and all
mathematical truth.  If Hilbert could do that, it would give us a mechanical
procedure to decide if a computer program will ever halt.  Why?

You just run through all possible proofs until you either find a proof
that the program halts or you find a proof that it never halts.
So if Hilbert's dream of a finite set of axioms
from which all of mathematical truth should follow
were possible, then by running through all possible proofs checking which
ones are correct, you would be able to decide if any computer program
halts.  In principle you could.  But
you {\bf can't} by Turing's very simple argument which is just Cantor's
diagonal argument applied to the computable reals.  That's how simple it is!

G\"odel's proof is ingenious and difficult.
Turing's argument is so fundamental, so deep, that everything seems natural and
inevitable.
But of course he's building on G\"odel's work.

\section*{3. The halting probability and algorithmic randomness}

The reason I talked to you about
Turing and computable reals is that I'm going to use a different procedure to
construct an uncomputable real, a much more uncomputable real than Turing does.
\[
\begin{array}{ll}
\begin{array}[t]{l}
   p_1 \;\; -\!.\underline{d_{11}}d_{12}d_{13}d_{14}d_{15}d_{16}\cdots \\
   p_2 \;\; -\!.d_{21}\underline{d_{22}}d_{23}d_{24}d_{25}d_{26}\cdots \\
   p_3 \;\; -\!.d_{31}d_{32}\underline{d_{33}}d_{34}d_{35}d_{36}\cdots \\
   p_4 \;\; -\!.d_{41}d_{42}d_{43}\underline{d_{44}}d_{45}d_{46}\cdots \\
   p_5 \;\;                                                            \\
   p_6 \;\; -\!.d_{61}d_{62}d_{63}d_{64}d_{65}\underline{d_{66}}\cdots \\
   \vdots                                                              \\
.\neq\!d_{11}\neq\!d_{22}\neq\!d_{33}\neq\!d_{44}\neq\!d_{55}\neq\!d_{66}\cdots
\end{array}
&
\begin{array}[t]{l}
   \mbox{{\bf G\"odel} 1931} \\
   \mbox{{\bf Turing}  1936} \\
   \mbox{uncomputable reals}
\end{array}
\end{array}
\]
And that's how we're going to get into much worse trouble.

How do I get
a much more uncomputable real?  (And I'll have to tell you how uncomputable it
is.)
Well, not with Cantor's diagonal argument.  I get this number, which I
like to call $\Omega$, like this:
\[
   \Omega = \sum_{\mbox{$p$ halts}} 2^{-|p|}
\]
This is just the halting probability.  It's sort of a mathematical pun.
Turing's fundamental result is that the halting problem is
unsolvable---there is no algorithm that'll settle the halting problem.
My fundamental result is that the halting probability is
algorithmically irreducible or algorithmically random.

What exactly is the halting
probability?   I've written down an expression for it:
\[
   \Omega = \sum_{\mbox{$p$ halts}} 2^{-|p|}
\]
Instead of looking at individual programs and asking whether they halt, you
put all computer programs together in a bag.  If you generate a computer
program at random by tossing a coin for each bit of the program,
what is the chance that the program will halt?
You're thinking of programs as bit strings, and you generate each bit by an
independent toss of a fair coin, so if a program is $N$ bits long, then the
probability that you get that particular program is $2^{-N}$.
Any program $p$ that halts contributes $2^{-|p|}$, two to the minus its
size in bits, the number of bits in it, to this halting probability.

By the way there's a technical detail which is very important and didn't work
in the
early version of algorithmic information theory.  You couldn't write this:
\[
   \Omega = \sum_{\mbox{$p$ halts}} 2^{-|p|}
\]
It would give infinity.  The technical detail is that no extension of a valid
program is a valid program.  Then this sum
\[
            \sum_{\mbox{$p$ halts}} 2^{-|p|}
\]
turns out to be between zero and one.  Otherwise it turns out to be infinity.
It only took ten years until I got it right.
The original 1960s version of algorithmic information theory
is wrong.  One of the reasons it's wrong is that you can't even define this
number
\[
   \Omega = \sum_{\mbox{$p$ halts}} 2^{-|p|}
\]
In 1974 I redid algorithmic information theory with ``self-delimiting''
programs
and then I discovered the halting probability $\Omega$.

Okay, so this is a probability
between zero and one
\[
   0 <
   \Omega = \sum_{\mbox{$p$ halts}} 2^{-|p|}
   < 1
\]
like all probabilities.  The idea is you generate each bit of a program
by tossing a coin and ask what is the probability that it halts.  This
number $\Omega$, this halting probability, is not only an uncomputable real---%
Turing already knew how to do that.  It is uncomputable in
the worst possible way.  Let me give you some clues how
uncomputable it is.

Well, one thing is it's algorithmically incompressible.
If you want to get the first $N$ bits of $\Omega$ out of a computer program,
if you want a computer program
that will print out the first $N$ bits of $\Omega$ and then halt,
that computer program has to be $N$ bits long.
Essentially you're only printing out constants that are in the program.
You cannot squeeze the first $N$ bits of $\Omega$.  This
\[
   0 <
   \Omega = \sum_{\mbox{$p$ halts}} 2^{-|p|}
   < 1
\]
is a real number, you could write it in binary.  And if you want to get
out the first $N$ bits from a computer program, essentially you just have
to put them in.  The program has to be $N$ bits long.  That's irreducible
algorithmic information.  There is no concise description.

Now that's an
abstract way of saying things.  Let me give a more concrete example of
how random $\Omega$ is.  \'Emile Borel at the turn of this century was one of
the
founders of probability theory.
\[
\begin{array}{l}
   \displaystyle
   0 <
   \Omega = \sum_{\mbox{$p$ halts}} 2^{-|p|}
   < 1
   \\
   \mbox{\bf \'Emile Borel}
\end{array}
\]

\begin{quotation}
{\sc Question:}  Can I ask a very simple question before you get ahead?

{\sc Answer:}  Sure.

{\sc Question:}  I can't see why $\Omega$ should be a probability.  What if
the two one-bit programs both halt?  I mean, what if the two one-bit programs
both halt and then some other program halts. Then $\Omega$ is greater than one
and not
a probability.

{\sc Answer:} I told you no extension of a valid program is a valid program.

{\sc Question:} Oh right, no other programs can halt.

{\sc Answer:}  The two one-bit programs would be all the programs there are.
That's the reason this number
\[
   0 <
   \Omega = \sum_{\mbox{$p$ halts}} 2^{-|p|}
   < 1
\]
can't be defined if you think of programs in the normal way.
\end{quotation}

So here we have \'Emile Borel, and he talked about something he called
a normal number.
\[
\begin{array}{l}
   \displaystyle
   0 <
   \Omega = \sum_{\mbox{$p$ halts}} 2^{-|p|}
   < 1
   \\
   \mbox{{\bf \'Emile Borel} --- normal reals}
\end{array}
\]
What is a normal real number?
People have calculated $\pi$ out to a billion digits,
maybe two billion.  One of the reasons for doing this,
besides that it's like climbing a mountain and having the world record,
is the question of whether each digit occurs the same number of
times.  It looks like the digits 0 through 9 each
occur 10\% of the time in the decimal expansion of $\pi$.
It looks that way, but nobody can prove it.  I think the same
is true for $\sqrt{2}$, although that's not as popular a number
to ask this about.

Let me describe some work Borel did
around the turn of the century when he
was pioneering modern probability theory.
Pick a real number in the unit interval,
a real number with a decimal point in front,
with no integer part.  If you pick a real number in the unit interval,
Borel showed that with probability one it's going to be ``normal.''  Normal
means that
when you write it in decimal each digit will occur in the limit exactly 10\%
of the time, and this will also happen in any other base.
For example in binary 0 and 1 will each occur in the
limit exactly 50\% of the time.  Similarly with blocks of digits.
This was called an absolutely normal real number
by Borel, and he showed that with probability one if you pick
a real number at random between zero and one it's going to have this property.
There's only one problem.  He didn't know whether $\pi$ is normal, he didn't
know whether $\sqrt{2}$ is normal.  In fact, he couldn't exhibit a single
individual example of a normal real number.

The first example
of a normal real number was discovered
by a friend of Alan Turing's at Cambridge called David Champernowne, who is
still alive and who's a well-known economist.  Turing was
impressed with him---I think he called him ``Champ''---because Champ had
published this in a paper as an undergraduate.
This number is known as Champernowne's number.  Let me show you
Champernowne's number.
\[
\begin{array}{l}
   \displaystyle
   0 <
   \Omega = \sum_{\mbox{$p$ halts}} 2^{-|p|}
   < 1
   \\
   \mbox{{\bf \'Emile Borel} --- normal reals}
   \\
   \mbox{\bf Champernowne}
   \\
   .01234567891011121314\cdots99100101\cdots
\end{array}
\]

It goes like this.  You write down a decimal point, then you write 0, 1, 2,
3, 4, 5, 6, 7, 8, 9, then 10, 11, 12, 13, 14 until 99, then 100, 101.
And you keep going in this funny way.
This is called Champernowne's number and Champernowne showed that it's
normal in base ten, only in base ten.  Nobody knows if it's normal
in other bases, I think it's still open.  In base ten though, not
only will the digits 0 through 9 occur exactly 10\% of the time in the limit,
but each possible block of two digits will
occur exactly 1\% of the time in the limit,
each block of three digits will occur exactly .1\% of the time in the limit,
etc.
That's called being normal in base ten.  But nobody knows what
happens in other bases.

The reason I'm saying all this is because it follows from the fact that the
halting probability $\Omega$ is
algorithmically irreducible information that this
\[
   0 <
   \Omega = \sum_{\mbox{$p$ halts}} 2^{-|p|}
   < 1
\]
is normal in any base.  That's easy to prove using
ideas about coding and compressing information that go back to Shannon.
So here we finally have an example of an absolutely normal number.  I don't
know how natural you
think it is, but it is a specific
real number that comes up and is normal in the most demanding
sense that Borel could think of.  Champernowne's number
couldn't quite do that.

This number $\Omega$ is in fact random in many
more senses.  I would say it this way.  It cannot be
distinguished from the result of independent tosses of a fair coin.
In fact this number
\[
   0 <
   \Omega = \sum_{\mbox{$p$ halts}} 2^{-|p|}
   < 1
\]
shows that you have total randomness and chaos and unpredictability
and lack of structure in pure mathematics!  The same way that all it
took for Turing to destroy Hilbert's dream was the diagonal argument,
you just write down this expression
\[
   0 <
   \Omega = \sum_{\mbox{$p$ halts}} 2^{-|p|}
   < 1
\]
and this shows that there are regions of pure mathematics
where reasoning is totally useless, where you're up against an impenetrable
wall.
This is all it takes.  It's just this halting probability.

Why do I say this?
Well, let's say you want to use axioms to prove what the bits of this
number $\Omega$ are.
I've already told you that it's uncomputable---right?---like the number that
Turing constructs using Cantor's diagonal argument.
So we know there is no algorithm which will compute digit by digit or bit by
bit
this number $\Omega$.  But let's try to prove what individual
bits are using a formal axiomatic system.
What happens?

The situation is very, very bad.  It's like this.
Suppose you have a formal axiomatic system which is $N$ bits of
formal axiomatic system (I'll explain what this means more precisely later).
It turns out that with a formal axiomatic system of complexity $N$, that is,
$N$ bits in size, you can prove what the positions and values are of at
most $N+c$ bits of $\Omega$.

Now what do I mean by formal axiomatic system
$N$ bits in size?  Well, remember that the essence of a formal axiomatic system
is
a mechanical procedure for checking whether a formal
proof follows the rules or not.  It's a computer program.  Of course in
Hilbert's days
there were no computer programs, but after Turing invented Turing machines
you could finally specify the notion of computer program
exactly, and of course now we're very familiar with it.

So the proof-checking algorithm which is the essence of any formal axiomatic
system in Hilbert's sense is a computer program,
and just see how many bits long this computer program is.\footnote
{Technical Note:  Actually, it's best to think of the complexity of a formal
axiomatic system as the size in bits of the computer program that enumerates
the set of all theorems.}
That's essentially how many bits it takes to specify the
rules of the game, the axioms and postulates and the rules of inference.
If that's $N$ bits, then you may be able to prove say that the first bit of
$\Omega$ in binary is 0, that the second bit is 1, that the third
bit is 0, and then there might be a gap, and you
might be able to prove that the thousandth bit is 1.  But you're only
going to be able to settle $N$ cases if your formal axiomatic system is an
$N$-bit formal axiomatic system.

Let me try to explain better what this means.
It means that you can only get out as much as you put in.
If you want to prove whether an individual bit in a
specific place in the binary expansion of the real number $\Omega$
is a 0 or a 1, essentially the only way to prove that is to take
it as a hypothesis, as an axiom, as a postulate.  It's irreducible
mathematical information.  That's the key phrase that really gives the
whole idea.
\[
\begin{array}{l}
   \mbox{\bf Irreducible Mathematical Information}
   \\
   \displaystyle
   0 <
   \Omega = \sum_{\mbox{$p$ halts}} 2^{-|p|}
   < 1
   \\
   \mbox{{\bf \'Emile Borel} --- normal reals}
   \\
   \mbox{\bf Champernowne}
   \\
   .01234567891011121314\cdots99100101\cdots
\end{array}
\]

Okay, so what have we got?
We have a rather simple mathematical object that
completely escapes us.  $\Omega$'s bits have no structure.  There is
no pattern, there is no structure that we as mathematicians can comprehend.
If you're interested in proving what individual bits of this number at specific
places are, whether they're 0 or 1, reasoning is completely useless.
Here mathematical reasoning is irrelevant and can get
nowhere.  As I said before, the only way a formal axiomatic system can get
out these results is essentially just to put them in as assumptions, which
means you're not using reasoning.  After all, anything can be
demonstrated by taking it as a postulate that you add to your set of axioms.
So this is a worst possible case---this is irreducible mathematical
information.  Here is a case where there is no structure, there are no
correlations, there is no pattern that we can perceive.

\section*{4. Randomness in arithmetic}

Okay, what does this have to do with randomness in arithmetic?
Now we're going back to G\"odel---I skipped over him rather quickly, and
now let's go back.

Turing says that you cannot use
proofs to decide whether a program will halt.  You can't always prove that
a program will halt or not.  That's how he destroys Hilbert's dream of
a universal mathematics.  I get us into more
trouble by looking at a different kind of question, namely, can you prove
that the fifth bit of this particular real number
\[
   0 <
   \Omega = \sum_{\mbox{$p$ halts}} 2^{-|p|}
   < 1
\]
is a 0 or a 1, or that the eighth bit is a 0 or a 1.  But these are
strange-looking questions.
Who had ever heard of the halting problem in 1936?
These are not the kind of things that mathematicians
normally worry about.  We're getting into trouble, but with
questions rather far removed from normal mathematics.

Even though you can't have a formal axiomatic system which can
always prove whether a program halts or not, it might be good for
everything else and then you could have an {\bf amended} version of
Hilbert's dream.  And the same with the halting probability
$\Omega$.  If the halting problem looks a little bizarre,
and it certainly did in 1936, well, $\Omega$ is brand new and certainly
looks bizarre.  Who ever heard of a halting probability?
It's not the kind of thing that mathematicians normally do.
So what do I care about all these incompleteness results!

Well, G\"odel had already faced this problem with his assertion which is true
but unprovable.
It's an assertion which says of itself that it's unprovable.
That kind of thing also never comes up in real mathematics.
One of the key elements in G\"odel's proof is that
he managed to construct an {\bf arithmetical} assertion
which says of itself that it's unprovable.
It was getting this self-referential assertion
to be in elementary number theory which took so much cleverness.

There's been a lot of work building on G\"odel's work,
showing that problems involving computations are
equivalent to arithmetical problems involving whole numbers.
A number of names come to mind.
Julia Robinson, Hilary Putnam and Martin Davis did some of the
important work, and then a key result was found in 1970 by Yuri
Matijasevi\v{c}.
He constructed a diophantine equation, which is an algebraic equation
involving only whole numbers, with a lot of variables.  One of the variables,
$K$,
is distinguished as a parameter.
It's a polynomial equation with integer coefficients and
all of the unknowns have to be whole numbers---that's a diophantine equation.
As I said, one of the unknowns is a parameter.
Matijasevi\v{c}'s equation has a solution for a particular value of the
parameter $K$
if and only if the $K$th computer program halts.

In the year 1900 Hilbert had asked
for an algorithm which will decide whether
a diophantine equation, an algebraic equation involving only whole numbers,
has a solution.
This was Hilbert's tenth problem. It was tenth is his famous list of
twenty-three problems.
What Matijasevi\v{c} showed in 1970 was that
this is equivalent to deciding whether an arbitrary computer
program halts.  So Turing's halting
problem is exactly as hard as Hilbert's tenth problem.  It's exactly
as hard to decide whether an arbitrary program will halt as to
decide whether an arbitrary algebraic equation in whole numbers has a
solution.
Therefore there is no
algorithm for doing that and Hilbert's tenth problem cannot be solved---%
that was Matijasevi\v{c}'s 1970 result.

Matijasevi\v{c} has gone on working in this area.
In particular there is a piece of work he did in collaboration with James Jones
in 1984.
I can use it to follow in G\"odel's footsteps, to follow G\"odel's example.
You see, I've shown that there's complete randomness, no pattern, lack of
structure,
and that reasoning is completely useless, if you're interested in the
individual
bits of this number
\[
   0 <
   \Omega = \sum_{\mbox{$p$ halts}} 2^{-|p|}
   < 1
\]

Following G\"odel, let's convert this into something in elementary number
theory.
Because if you can get into all this trouble
in elementary number theory, that's the bedrock.
Elementary number theory,
1, 2, 3, 4, 5, addition and multiplication, that goes back to the
ancient Greeks and it's the most solid part of
all of mathematics.  In set theory you're
dealing with strange objects like large cardinals, but here you're not even
dealing with
derivatives or integrals or measure, only with whole numbers.  And using the
1984 results of
Jones and Matijasevi\v{c} I can indeed dress up $\Omega$ arithmetically and get
randomness in elementary number theory.

What I get is an
exponential diophantine equation with a parameter.  ``Exponential
diophantine equation'' just means that you allow variables in the
exponents. In contrast, what Matijasevi\v{c} used to show that Hilbert's tenth
problem is unsolvable is just a polynomial diophantine
equation, which means that the exponents are always natural number constants.
I have to allow $X^Y$.  It's not known yet whether I actually
need to do this.  It might be the case that I can manage with a
polynomial diophantine equation.  It's an open question, I believe
that it's not settled yet.  But for now, what I have is an exponential
diophantine equation with seventeen thousand variables.  This equation is
two-hundred pages long and again one variable is the parameter.

This is an equation where every constant is a whole number, a natural number,
and all the variables are also natural numbers, that is,
positive integers.  (Actually {\bf non-negative} integers.)
One of the variables is a parameter, and you change the value
of this parameter---take it to be 1, 2, 3, 4, 5.  Then
you ask, does the equation have a finite or infinite
number of solutions?  My equation is constructed so that
it has a finite number of solutions if a particular individual
bit of $\Omega$ is a 0, and it has an infinite number of solutions if
that bit is a 1.  So deciding whether my exponential diophantine equation
in each individual case has a finite or infinite number of solutions
is exactly the same as determining what an individual bit of this
\[
   0 <
   \Omega = \sum_{\mbox{$p$ halts}} 2^{-|p|}
   < 1
\]
halting probability is.  And this is completely intractable
because $\Omega$ is irreducible mathematical information.

Let me emphasize the difference between this and Matijasevi\v{c}'s work on
Hilbert's
tenth problem.  Matijasevi\v{c} showed that there is a
polynomial diophantine equation with a parameter with the following property:
You vary
the parameter and ask, does the equation have a solution?  That turns out to be
equivalent to Turing's halting problem, and therefore escapes the power of
mathematical reasoning,
of formal axiomatic reasoning.

How does this differ from what I do?
I use an exponential diophantine equation, which means I allow variables in the
exponent.
Matijasevi\v{c} only allows constant exponents.
The big difference is that Hilbert asked for an algorithm to
decide if a diophantine equation has a solution.
The question I have to ask to get randomness in elementary number theory,
in the arithmetic of the natural numbers, is slightly more sophisticated.
Instead of asking whether there is a solution,
I ask whether there are a finite or infinite number of solutions---a more
abstract question.
This difference is necessary.

My two-hundred page equation is constructed so that it has a finite or
infinite number of solutions depending on whether a particular bit of
the halting probability is a 0 or a 1.
As you vary the parameter, you get each individual bit of $\Omega$.
Matijasevi\v{c}'s equation is constructed so that it has a solution
if and only if a particular program ever halts.
As you vary the parameter, you get each individual computer program.

Thus even in arithmetic you can find $\Omega$'s absolute lack of structure,
$\Omega$'s randomness and irreducible mathematical information.
Reasoning is completely powerless in those areas of arithmetic.
My equation shows that this is so.
As I said before, to get this equation I use ideas that start in G\"odel's
original 1931 paper.
But it was Jones and Matijasevi\v{c}'s 1984 paper that finally gave me the tool
that I needed.

So that's why I say that there is randomness in elementary number theory,
in the arithmetic of the natural numbers.
This is an impenetrable stone wall, it's a worst case.
{}From G\"odel we knew that we couldn't get a formal axiomatic system to be
complete.
We knew we were in trouble, and Turing showed us how basic it was,
but $\Omega$ is an extreme case where reasoning fails completely.

I won't go into the details, but let me talk in vague information-theoretic
terms.
Matijasevi\v{c}'s equation gives you $N$ arithmetical questions with yes/no
answers
which turn out to be only $\log N$ bits of algorithmic information.
My equation gives you $N$ arithmetical questions with yes/no answers
which are irreducible, incompressible mathematical information.

\section*{5. Experimental mathematics}

Okay, let me say a little bit in the minutes I have left about what this all
means.

First of all, the connection with physics.
There was a big controversy when quantum mechanics was developed,
because quantum theory is nondeterministic.  Einstein didn't like that.  He
said,
``God doesn't play dice!''  But as I'm sure you all know, with chaos and
nonlinear dynamics we've now realized that even in classical physics we
get randomness and unpredictability.  My work is in the
same spirit.  It shows that pure mathematics, in fact even
elementary number theory, the arithmetic of the natural numbers, 1, 2, 3,
4, 5, is in the same boat.  We get randomness there too.
So, as a newspaper headline would put it,
God not only plays dice in quantum mechanics
and in classical physics, but even in pure mathematics,
even in elementary number theory.  So if a new
paradigm is emerging, randomness is at the heart of it.
By the way, randomness is also at the heart of quantum field theory,
as virtual particles and Feynman path integrals (sums over all histories) show
very clearly.
So my work fits in with a lot of work in physics, which is why I often get
invited
to talk at physics meetings.

However the really important question isn't physics, it's mathematics.
I've heard that G\"odel wrote a letter to his mother who stayed in Europe.
You know, G\"odel and Einstein were friends at the Institute for Advanced
Study.
You'd see them walking down the street together.
Apparently G\"odel
wrote a letter to his mother saying that even though Einstein's work on
physics had really had a tremendous impact on how people did physics,
he was disappointed that his work had not had the same effect on
mathematicians.
It hadn't made a difference
in how mathematicians actually carried on their everyday work.  So I
think that's the key question:  How should you really do mathematics?

I'm claiming I have a much stronger incompleteness result.  If so maybe it'll
be clearer whether mathematics should be done the ordinary way.  What is the
ordinary
way of doing mathematics?  In spite of the fact that everyone knows
that any finite set of axioms is incomplete,
how do mathematicians actually work?    Well suppose you have a conjecture that
you've been
thinking about for a few weeks, and you believe it because you've
tested a large number of cases on a computer.  Maybe it's a conjecture about
the primes and
for two weeks you've tried to prove it.  At the end of two weeks you don't say,
well obviously the reason I haven't been able to show this is because of
G\"odel's incompleteness theorem!  Let us therefore add it
as a new axiom!  But if you took G\"odel's incompleteness theorem
very seriously this might in fact be the way to proceed.  Mathematicians will
laugh
but physicists actually behave this way.

Look at the history of physics.
You start with Newtonian physics.  You cannot get Maxwell's equations from
Newtonian physics.  It's a new domain of experience---you need new postulates
to deal with it.  As for special relativity, well, special relativity is almost
in Maxwell's equations.  But Schr\"odinger's equation does not come from
Newtonian physics and Maxwell's equations.  It's a new domain of experience
and again you need new axioms.  So physicists are used to the idea that when
you start experimenting at a smaller scale, or with new phenomena,
you may need new principles to understand and explain what's going on.

Now in spite of incompleteness mathematicians don't behave at all like
physicists do.
At a subconscious level they still assume that the small
number of principles, of postulates and methods of inference,
that they learned early as mathematics students, are enough.
In their hearts they believe
that if you can't prove a result it's your own fault.  That's
probably a good attitude to take rather than to blame someone else, but let's
look at a question like the Riemann hypothesis.
A physicist would say that there is ample experimental evidence for the Riemann
hypothesis
and would go ahead and take it as a working assumption.

What is the Riemann hypothesis?  There are many unsolved questions involving
the distribution
of the prime numbers that can be settled if you assume the Riemann hypothesis.
Using computers people check these conjectures and they work
beautifully.  They're neat formulas but nobody can prove them.
A lot of them follow from the Riemann hypothesis.  To a physicist this would
be enough:  It's useful, it explains a lot of data.  Of course a physicist then
has
to be prepared to say ``Oh oh, I goofed!''\ because an experiment can
subsequently
contradict a theory.  This happens very often.

In particle physics you throw up theories all the time and most of them
quickly die.  But mathematicians don't like to have to backpedal.
But if you play it safe,
the problem is that you may be losing out, and I believe you are.

I think it should be obvious where I'm leading.
I believe that elementary number theory and the rest of mathematics
should be pursued more in the spirit of experimental science, and that you
should be willing to adopt new principles.  I believe that Euclid's statement
that an axiom is a self-evident truth is a big mistake.
The Schr\"odinger equation certainly isn't a self-evident truth!  And the
Riemann hypothesis isn't self-evident either, but it's very useful.

So I believe that we mathematicians shouldn't ignore incompleteness.  It's a
safe thing to
do but we're losing out on results that we could get.  It would be as if
physicists said, okay no Schr\"odinger equation, no Maxwell's equations, we
stick with Newton, everything must be deduced from Newton's laws.
(Maxwell even tried it.  He had a mechanical model of an electromagnetic
field.  Fortunately they don't teach that in college!)

I proposed all this twenty years ago
when I started getting these information-theoretic incompleteness results.
But independently a new school on the philosophy of
mathematics is emerging called the ``quasi-empirical'' school of thought
regarding the foundations
of mathematics.  There's a book of Tymoczko's called
{\it New Directions in the Philosophy of Mathematics\/}
(Birkh\"auser, Boston, 1986).  It's a good collection of articles.
Another place to look is {\it Searching for Certainty\/} by John Casti
(Morrow, New York, 1990) which has a good chapter on mathematics.  The last
half of the chapter talks about this quasi-empirical view.

By the way, Lakatos, who was one of the people involved in this new movement,
happened to be at Cambridge at that time.  He'd left Hungary.

The main schools of mathematical philosophy
at the beginning of this century were Russell and Whitehead's
view that logic was the basis for everything, the formalist school of Hilbert,
and an ``intuitionist'' constructivist school of Brouwer.
Some people think that Hilbert believed that
mathematics is a meaningless game played with marks of ink on paper.
Not so!
He just said that to be absolutely clear and precise what mathematics is all
about, we have to specify the
rules determining whether a proof is correct so precisely that they become
mechanical.
Nobody who thought that mathematics
is meaningless would have been so energetic and done such important
work and been such an inspiring leader.

Originally
most mathematicians backed Hilbert. Even after G\"odel
and even more emphatically Turing showed that
Hilbert's dream
didn't work, in practice mathematicians carried on as before, in Hilbert's
spirit.
Brouwer's constructivist attitude was mostly considered a nuisance.
As for Russell and Whitehead, they had a lot of
problems getting all of mathematics from logic.  If you
get all of mathematics from set theory you discover that it's nice to define
the whole numbers in terms of sets (von Neumann worked on this).
But then it turns out that there's all kinds of problems with sets.
You're not making the natural numbers more solid by basing them on something
which is
more problematical.

Now everything has gone
topsy-turvy.  It's gone topsy-turvy, not because of any philosophical
argument, not because of G\"odel's
results or Turing's results or my own incompleteness results.
It's gone topsy-turvy for a very simple reason---the computer!

The computer as you all know has changed the way we do everything.
The computer has
enormously and vastly increased mathematical experience.  It's so
easy to do calculations,
to test many cases, to run experiments on the computer.
The computer has
so vastly increased mathematical experience, that in order to cope,
people are forced to proceed in a more pragmatic
fashion.
Mathematicians are proceeding more pragmatically, more
like experimental scientists do.
This new tendency is often called ``experimental mathematics.''
This phrase comes up a lot in the field of chaos, fractals and nonlinear
dynamics.

It's often the case that when doing experiments on the computer,
numerical experiments with equations,
you see that something happens, and you conjecture a result.
Of course it's nice if you can prove it.  Especially if the proof is short.
I'm not sure that a thousand
page proof helps too much.  But if it's a short proof it's
certainly better than not having a proof.  And if you have several proofs from
different viewpoints, that's very good.

But sometimes
you can't find a proof and you can't wait for
someone else to find a proof, and you've got to
carry on as best you can.  So now mathematicians sometimes
go ahead with working hypotheses on the basis of the results
of computer experiments.  Of course if it's physicists doing these computer
experiments, then it's certainly okay; they've always relied heavily on
experiments.
But now even mathematicians sometimes operate in this manner.
I believe that there's a new journal called the {\it Journal of Experimental
Mathematics.}  They should've put me on their editorial board, because
I've been proposing this for twenty years based on my
information-theoretic ideas.

So in the end it wasn't G\"odel,
it wasn't Turing, and it wasn't my results that are making mathematics go in an
experimental mathematics direction, in a quasi-empirical direction.  The reason
that mathematicians are changing their working habits is the computer.
I think it's an excellent joke!
(It's also funny that of the three old schools of mathematical philosophy,
logicist, formalist, and intuitionist, the most neglected was Brouwer,
who had a constructivist attitude years before the computer gave a tremendous
impulse to constructivism.)

Of course, the mere fact that everybody's doing something doesn't mean that
they ought to be.
The change in how people are behaving isn't because of
G\"odel's theorem or Turing's theorems or my theorems, it's because of the
computer.  But I think that the sequence of work that I've outlined does
provide some theoretical justification for what everybody's
doing anyway without worrying about the theoretical justification.  And I think
that the question
of how we should actually do mathematics requires {\bf at least} another
generation of work.
That's basically what I wanted to say---thank you very much!

\section*{Bibliography}

\begin{itemize}
\item[{[1]}] G. J. Chaitin, {\it Information-Theoretic Incompleteness,}
World Scientific, 1992.
\item[{[2]}] G. J. Chaitin, {\it Information, Randomness \& Incompleteness,}
second edition,
World Scientific, 1990.
\item[{[3]}] G. J. Chaitin, {\it Algorithmic Information Theory,} revised third
printing,
Cambridge University Press, 1990.
\end{itemize}

\end{document}